\documentclass[aip,cha,amsmath,amssymb,reprint,numerical]{revtex4-1}

\usepackage{graphicx}
\usepackage{dcolumn}
\usepackage{bm}
\usepackage{subfigure}
\usepackage{color}
\usepackage{verbatim}
\usepackage{epsfig}

\begin{document}


\title{Spatio-temporal organization of dynamics in a two-dimensional periodically driven vortex flow: a Lagrangian flow network perspective}%

\author{Michael Lindner}
\affiliation{Potsdam Institute for Climate Impact Research, Telegrafenberg A31, 14473 Potsdam, Germany}
\affiliation{Department of Mathematics, Humboldt University, Rudower Chaussee 25, 12489 Berlin,
Germany}

\author{Reik V. Donner}%
\affiliation{Potsdam Institute for Climate Impact Research, Telegrafenberg A31, 14473 Potsdam, Germany}

\date{\today}

\begin{abstract}

We study the Lagrangian dynamics of passive tracers in a simple model of a driven two-dimensional vortex resembling real-world geophysical flow patterns. Using a discrete approximation of the system's transfer operator, we construct a directed network that characterizes the exchange of mass between distinct regions of the flow domain. By studying different measures characterizing flow network connectivity at different time-scales, we are able to identify the location of dynamically invariant structures and regions of maximum dispersion. Specifically, our approach allows to delimit co-existing flow regimes with different dynamics. To validate our findings, we compare several network characteristics to the well-established finite-time Lyapunov exponents and apply a receiver operating characteristic (ROC) analysis to identify network measures that are particularly useful for unveiling the skeleton of Lagrangian chaos.

\end{abstract}


\maketitle

\begin{quotation}

Complex networks provide a timely tool for investigating structural properties associated with the organization of dynamical processes in ``spatially extended'' systems. Among others, flow patterns (in both general continuous-time dynamical systems and real-world geophysical fluid dynamics contexts) can be conveniently described by directed network representations. In this case, nodes denote certain parts of the domain of interest in phase or physical space, and weighted links represent the probability that when tracing the system's evolution starting within one region, after a given time one may observe a state in the other region. This work further explores the potentials of some complex network measures for characterizing the Lagrangian dynamics of the flow based on passively advected tracers. By constructing such Lagrangian flow networks for a paradigmatic model system exhibiting a classical vortex pattern common to many real-world geophysical flows, the potentials of this approach for characterizing mixing and dispersion within the flow domain are demonstrated. Specifically, a suite of complementary network measures allows a systematic partitioning of the flow into domains with different dynamical characteristics.

\end{quotation}

\section{\label{sec:intro}Introduction}

During the last decades, complex network theory has been developed as a new powerful paradigm of nonlinear sciences~\cite{Albert2002,Newman2010}. Recently, there has been a growing body of literature on applying this framework to geophysical problems, including (among others) network representations of the Earth's climate system based on the statistical similarity between the climate dynamics observed at different points on the Earth~\cite{Tsonis2004,Radebach2013}. Besides the consideration of scalar fields (like temperature or pressure) commonly used in such climate networks, recent work has been devoted to identifying dynamically relevant structures from network representations of vector fields based on correlations between the local magnitudes of such vector fields~\cite{Molkenthin2013} or canonical correlations~\cite{Dods2015}. 

As an alternative to these similarity-based approaches, the passive advection of particles -- allowing to estimate transition probabilities between different parts of the spatial domain of interest -- has been proposed as a promising tool for studying dynamical patterns in geophysical flows~\cite{rossi2014hydrodynamic,sergiacomi2015flow}. Conceptually, this approach is closely related to the idea of transition networks based on symbolic dynamics representations of a trajectory (or finite time series) of a dynamical system, where a discretized version of the system's state space is considered in terms of transition probabilities between such discrete states~\cite{Nicolis2005,Padberg2009,Campanharo2011,Small2013,McCullough2015}. Another notable and closely related approach has been recently proposed by Hadjighasem \textit{et~al.}~\cite{Hadjighasem2015}, who considered Lagrangian trajectories with different initial conditions and used weighted graph representations based on the mutual distances between such trajectories to obtain a clustering of phase space for detecting coherent vortices embedded in the flow.

This work is dedicated to further study the potentials of complex network approaches to investigating flow structures. In general, having a flow as described by a velocity field, fluid dynamics distinguishes two different perspectives: On the one hand, the ``direct'' (Eulerian) view focuses on the structures exhibited by the velocity field itself. In the context of recent work, we mention the flow network approach by Molkenthin~\textit{et~al.}~\cite{Molkenthin2013} as an example for the application of complex network tools. On the other hand, as mentioned above, Lagrangian dynamics approaches studying the trajectories of passive particles advected within the flow allow obtaining a different type of network representation. Notably, Eulerian dynamics (\emph{of} the velocity field) and Lagrangian dynamics (tracer advection \emph{in} the field) reveal complementary information on the flow pattern under study. For example, velocity fields with regular time evolution (e.g., periodic two-dimensional or even stationary three-dimensional flows) may result in chaotic particle trajectories (Lagrangian chaos).

In the remainder of this work, we exclusively focus on the Lagrangian description of the flow, which has proven especially useful for studying transport and mixing phenomena. The most prominent methods in this field can be classified into \emph{geometric} and \emph{probabilistic} approaches. Among other techniques, finite-time Lyapunov exponents (FTLE) and related stretching indicators have been widely used to identify Lagrangian coherent structures (LCS)~\cite{haller2000lagrangian, haller2001distinguished, haller2011variational, shadden2005definition}, i.e., quasi-stationary patterns of spatially confined particle dynamics. These structures are directly related with geometric objects like material lines or surfaces that persist for relatively long times and thus constitute barriers to transport. 

The probabilistic approaches of Lagrangian dynamics build upon approximations of the transfer operator (Perron-Frobenius operator) of the system under study. Here, instead of studying transport barriers these methods aim at directly identifying regions that stay (almost) invariant under the dynamics~\cite{froyland2009almost, froyland2010transport, froyland2013analytic}. The main tool of these approaches is the transition matrix, a discrete approximation of the transfer operator, which indicates the amount of flow between different subsets of the flow domain during a certain time interval. The analysis of the transition matrix has already been demonstrated to reveal interesting properties of the flow: By studying its eigenvectors, Froyland \textit{et al.}\, identified almost-invariant sets in time-dependent flows~\cite{froyland2009almost} and demonstrated how to detect the approximate locations of the five so-called \emph{garbage patches} in the global ocean surface flow~\cite{froyland2014well}. These patches were also identified by van Sebille \textit{et al.}~\cite{vansebille2012origin}, who constructed the transition matrix from data on observed surface drifters. Extending upon these ideas, Froyland \textit{et al.} proposed a novel measure of nonlinear stretching based on the transition matrix, the so-called 
\emph{finite-time entropy}~\cite{froyland2012finite}, which has been generalized recently to a family of (R\'enyi-like) network entropies~\cite{sergiacomi2015flow}.

Some recent approaches have interpreted the transition matrix as the adjacency matrix of a weighted and directed network and thereby made the rich toolbox of complex network techniques available for flow system analysis. Ser-Giacomi \textit{et al.} related the node degree to nonlinear stretching~\cite{sergiacomi2015flow} and obtained maximum flow pathways from betweenness centrality~\cite{sergiacomi2014most}. To discern hydrodynamic provinces in the Mediterranean, Rossi \textit{et al.}~\cite{rossi2014hydrodynamic} identified communities in the associated flow network.

This paper aims to further contribute to a better understanding of different topological characteristics of such Lagrangian flow networks and their relation to the underlying flow. For this purpose, we consider a suite of established complex network measures as well as related characteristics accounting for the finite transport range of the flow within a given finite period of time. Subsequently, these measures are applied to a well understood model of a two-dimensional driven vortex flow~\cite{feudel2005intersections}. In order to test the ability of different network measures to distinguish between regions of different dispersion and mixing properties, we use the corresponding FTLE field as a reference and compare the considered flow network characteristics by means of a receiver-operating characteristic (ROC) analysis~\cite{fawcett2006introduction}, a tool widely used in statistics and machine learning. 

Our results demonstrate that not only the node degree, but also other conceptually related node characteristics of Lagrangian transport networks are tightly connected with the FTLE field and can thus be used to obtain similar information on the flow pattern under study. More specifically, we show that regions of laminar and strongly mixing dynamics can be clearly distinguished in terms of closeness, and that the finite-time structures associated with invariant manifolds embedded in the flow can be well approximated at different temporal scales using the newly defined measure cutoff closeness.

The remainder of this paper is organized as follows: In Section~\ref{sec:methods}, the theoretical foundations of our study are reviewed, including the definition of FTLEs and the construction of the transition matrix. Subsequently, Section~\ref{sec:network} discusses the concept of Lagrangian flow networks together with a selection of network measures and introduces a new time-sensitive extension of path-based measures by making use of a cutoff, which is further detailed taking closeness as a particular example. In Section~\ref{sec:example}, we briefly present the driven vortex system as a test case for our approach. The results obtained for this system using Lagrangian flow networks are presented and further discussed in Section \ref{sec:results}. Finally, Section~\ref{sec:discussion} summarizes the findings of this work and briefly discusses possible directions for further studies.

\section{\label{sec:methods}Methodological background}

\subsection{Flow field and flow map}

We consider the velocity of particles in a flow given by an ordinary differential equation~\cite{arnold1992}
\begin{equation}
\frac{d}{dt}\vec{x}\left(t\right)=\vec{F}\left(\vec{x}\left(t\right),t\right).\label{eq:ode}
\end{equation}

\noindent
We require $\vec{F}:\Omega\times\mathbb{R}\rightarrow\Omega$ to be a continuously differentiable vector field defined on a smooth, compact manifold $\Omega\subset\mathbb{R}^{n}$. If $\vec{F}\left(\vec{x},t\right)=\vec{F}\left(\vec{x}\right)$ the system is called \emph{autonomous}, otherwise \emph{non-autonomous} or \emph{time-dependent}. The associated \emph{flow map} 
\[
\Phi:\Omega\times\mathbb{R}\times\mathbb{R}\rightarrow\Omega;\,\left(\vec{x}_0 , t_0,\tau\right)\mapsto\vec{x}\left(t_{0}+\tau\right)
\]
gives the final position of a particle released at point $\vec{x}_0$ at time $t_0$
after being passively advected for a time $\tau$. Here, $\vec{x}:\mathbb{R}\mapsto\Omega$ denotes the solution to Eq.~\eqref{eq:ode} with initial condition  ${\vec{x}\left(t_0\right)=\vec{x}_0}$.

Because of our requirements on $\Omega$ and $\vec{F}$, standard theorems on the local
existence and uniqueness of solutions of \eqref{eq:ode} ensure, that
the flow map is continuous and satisfies the following properties:
\begin{eqnarray}
\Phi\left(\vec{x}_0,t_{0},0\right) & = & \vec{x}\left(t_{0}\right); \nonumber \\
\Phi\left(\Phi\left(\vec{x}_0,t_{0},\tau_{1}\right),t_{0}+\tau_{1},\tau_{2}\right) 
& = & 
\Phi\left(\vec{x}\left(t_{0}+\tau_{1}\right),t_{0}+\tau_{1},\tau_{2}\right) 
\nonumber \\
 & = & 
\Phi\left(\vec{x}\left(t_{0}+\tau_{2}\right),t_{0}+\tau_{2},\tau_{1}
\right) \nonumber \\
 & = & \vec{x}\left(t_{0}+\tau_{1}+\tau_{2}\right).
\end{eqnarray}
\noindent
Therefore, $\Phi$ defines a discrete dynamical system. Note that the above requirements ensure that
\begin{equation}
 \Phi^{-1}\left(\vec{x}\left(t_0\right),t_{0},\tau\right) = \Phi\left(\vec{x}\left(t_0+\tau\right),t_{0}+\tau,-\tau\right).
\end{equation}

\subsection{Finite-time Lyapunov exponents}

In dynamical systems approaches to flow processes, a standard way to quantify dispersion and mixing in fluids is based on finite-time Lyapunov exponents (FTLEs)~\cite{shadden2005definition}.
The FTLE $\lambda\left(\vec{x}_{0},t_{0},\tau\right)$ characterizes the amount of stretching about the trajectory starting at point $\vec{x}_{0}$ in the time interval $\left[t_{0},t_{0}+\tau\right]$ and is defined as
\begin{equation}
\lambda\left(\vec{x}_{0},t_{0},\tau\right)=\frac{1}{2\left|\tau\right|}
\log\Lambda_{max},\label{eq:ftle}
\end{equation}
\noindent
where $\Lambda_{max}$ is the largest eigenvalue of the Cauchy-Green
strain tensor
\begin{equation}\label{eq:cauchygreen}
C\left(\vec{x}_{0},t_{0},\tau\right)=\left(\nabla\Phi\left(\vec{x}_{0},t_{0},
\tau\right)\right)^{T}\nabla\Phi\left(\vec{x}_{0},t_{0},\tau\right).
\end{equation}
\noindent
Here, $\nabla\Phi\left(\vec{x}_{0},t_{0},\tau\right)$ is the Jacobian matrix of the flow map~\cite{sergiacomi2015flow, shadden2005definition}. For $\tau>0$, $\lambda$ is referred to as the forward FTLE and characterizes dispersion around $x_{0}$. For $\tau<0$, we obtain the backward FTLE related to the strength of mixing around $x_{0}$.

We recall that the interpretation of Eq.~\eqref{eq:ftle} is that an initial sphere of infinitesimal diameter $r$, located at $\vec{x}_{0}$, will be stretched into an ellipsoid of major axis $r\exp\left(\tau\cdot\lambda\left(\vec{x}_{0},t_{0},\tau\right)\right)$ after being advected by the flow for a time $\tau$. Similarly, the other axes will be deformed at exponential rate related to the other eigenvalues of Eq.~\eqref{eq:cauchygreen}. Thus, for high values of the forward FTLE $\lambda\left(\vec{x}_{0},t_{0},\tau\right)$ the trajectories of two particles that are very close to $\vec{x}_{0}$ at $t_{0}$ will separate at exponential rate, ending up in different regions of the flow domain. Consequently, high forward FTLE values correspond to regions with strong dispersion of particles. Conversely, a high backward FTLE at $x_{0}$ implies that trajectories that were far apart at time $t_{0}-\tau$, become very close at $t_{0}$, which corresponds to an efficient mixing of fluid from different regions at $x_{0}$.

For example, in the case of a hyperbolic fixed point embedded in a steady flow, one expects high forward FTLE near the stable manifold and high backward FTLE near the unstable manifold. These manifolds act as \emph{separatrices} or transport barriers, because they separate regions with different long-term dynamics. For time-dependent systems, the notions of saddle points and their stable and unstable manifolds are not well-defined anymore. However, it is often possible to interpret so-called \emph{Lagrangian Coherent
Structures}~\cite{haller2000lagrangian, haller2001distinguished, haller2011variational} in terms of finite-time separatrices associated with time-dependent hyperbolic objects embedded within the flow. Shadden \textit{et~al.}~\cite{shadden2005definition} demonstrated how to identify LCSs from the ridges of the FTLE field and that in many situations the flux across such ridges is small or even negligible.

\subsection{Transfer operator and transition matrix}

In the context of numerical computations to study flow properties, it is usually necessary to discretize the spatial domain of the flow. In this work, we partition the flow domain $\Omega$ into a sufficiently large number $N$ of spatially connected yet mutually disjoint boxes $B_{i}$, $i=1,\dots,N$. For convenience, we consider these boxes to be of equal volume. Note that such a partition always exists since $\Omega$ is required to be compact.

The proportion of mass that flows from box $B_{i}$ to box $B_{j}$ within the time interval $\left[t_{0},t_{0}+\tau\right]$ is given by
\begin{equation}
P_{ij}\left(t_{0},\tau\right)=\frac{\text{vol}\left(B_{i}\cap\Phi\left(B_
{j},t_{0}+\tau, -\tau\right)\right)}{\text{vol}\left(B_{i}\right)}.
\label{eq:ulam}
\end{equation}
\noindent
Here, $\text{vol}\left(X\right)$ denotes the Lebesque measure of a subset $X\in\Omega\subset\mathbb{R}^{n}$. Equation~\eqref{eq:ulam} defines the \emph{transition matrix} $\mathbf{P}\left(t_{0},\tau\right)$ of the flow, which provides a spatially discretized approximation to the \emph{transfer operator} or \emph{Perron-Frobenius operator} of the dynamical system. This construction is commonly known as \emph{Ulam's method}~\cite{froyland2009almost,froyland2001extracting}.

\subsection{Numerical approximation of the transition matrix}

To obtain a numerical estimate $\hat{\mathbf{P}}\left(t_{0},\tau\right)$ of the transition matrix, at time $t_{0}$ we initialize a large number $K_i$ of particles within each box $B_{i}$ at positions $\vec{x}_{i,k}$ ($k=1,\dots,K_i$) uniformly at random and numerically integrate the corresponding trajectory segments to obtain $\Phi\left(\vec{x}_{i,k},t_{0},\tau\right)$. The number of particles transported from box $B_{i}$ to $B_{j}$ provides an estimate of the flow between the two boxes, i.e.,
\begin{equation}
\hat{P}_{ij}\left(t_{0},\tau\right)=\frac{\#\left\{ k:x_{i,k}\in 
B_{i}\wedge\Phi\left(x_{i,k},t_{0},\tau\right)\in B_{j}\right\} 
}{K_i}.\label{eq:numericUlam}
\end{equation}
\noindent
Here, $\#X$ is the number of elements of a finite set $X$. The integration time $\tau$ should be chosen long enough, so that sufficiently many particles have left their initial box. Otherwise, the approximated transport matrix will be close to the identity matrix. There is no upper limit on $\tau$, but it should be chosen of appropriate size to resolve the particular time-scales of interest of the flow. Obviously the latter applies to the choice of $t_{0}$ as well.

To ensure that no significant part of the dynamics is neglected in boxes with 
high particle dispersion, we adaptively increase $K_i$ until 
\begin{equation}
  \sum_j \hat{P}_{ij}\left(t_0,\tau\right) < \alpha K_i  \label{eq:adap}
\end{equation}
\noindent
with $\alpha=0.1$. This condition ensures that the number of initial conditions per box is significantly higher than the number of boxes reached after integration. While we observe that this leaves many resulting flow network characteristics (see below) practically unchanged, it strongly reduces computation time, since the number of boxes reached within a time step $\tau$ usually differs by several orders of magnitude between regions of chaotic and laminar dynamics. It has to be noted that the adaptive method slightly favors out-degree over in-degree (see Section~\ref{sec:network} for the definition of these measures), leading to about 10 to 20 percent higher maximum out-degree than maximum in-degree as will be shown in Section~\ref{sec:results}. However, at this point we accept this potential drawback since the studied network measures are mostly related to the forward dynamics (e.g., out-degree). If being interested in the backward dynamics, either a uniform $K_i$ should be used or the integration could be performed backward with adaptive $K_i$.

\subsection{Markov chain description}

A \emph{Markov chain}~\cite{behrends2000introduction} is a memoryless stochastic process, which means that the state of the process after $k$ time steps depends
only on its state at time step $k-1$.

For our purposes, we will consider an $N$-state Markov chain. Let $\rho_{i}\left(t_{0}\right)$ be the probability of a particle to be in box $B_{i}$ at time $t_{0}$ and consider the vector of residence probabilities $\vec{\rho}\left(t_{0}\right)=\left(\rho_{1}\left(t_{0}\right),\rho_{2}\left(t_{0}\right),\dots,\rho_{N}\left(t_{0}\right)\right)$. The evolution of this vector during a discrete time step $\tau$ is given by
(left-) multiplication with the transition matrix,
\begin{equation}
\vec{\rho}\left(t_{0}+\tau\right)=\vec{\rho}\left(t_{0}\right)\mathbf{P}\left(t_{0}
,\tau\right).
\end{equation}
In this case, $P_{ij}\left(t_{0},\tau\right)$ gives the conditional transition probability
between boxes, that is, the probability that a particle ends up in box $B_{j}$ after time step $\tau$, given that it was initially in box $B_{i}$. Thus, the transition matrix $\mathbf{P}\left(t_{0},\tau\right)$ approximates the forward dynamics of the flow. The corresponding backward evolution is given by $\mathbf{P}\left(t_{0}+\tau,-\tau\right)$, which can be easily obtained from $\mathbf{P}\left(t_{0},\tau\right)$ via~\cite{froyland2012finite}
\begin{equation}
P_{ji}\left(t_{0}+\tau,-\tau\right)=\frac{P_{ij}\left(t_{0},\tau\right)}{\sum_{k=1}^{N} P_{kj}\left(t_{0},\tau\right)}.\label{eq:backwardP}
\end{equation}
\noindent
Equations~\eqref{eq:ulam} and \eqref{eq:numericUlam} ensure that the matrix $\mathbf{P}\left(t_{0},\tau\right)$ is \emph{row-stochastic}, that is, $\sum_{j=1}^{N} P_{ij}\left(t_{0},\tau\right)=1$ for
all $i$ and $P_{ij}\left(t_{0},\tau\right)\geq 0$ for all $i,j$. Therefore, according to standard theorems on stochastic matrices, $\mathbf{P}\left(t_{0},\tau\right)$ has an eigenvalue equal to $1$ as its the largest eigenvalue, and at least one corresponding left eigenvector has only non-negative elements~\cite{froyland2009almost}. Froyland~\textit{et~al.}\cite{froyland2014well} used eigenvectors of the transition matrix with eigenvalues close to 1 to identify attracting
sets and their basins of attraction in the global ocean surface flow.

Since Markov processes are memoryless, this framework allows approximating a long-term transition matrix 
$\mathbf{P}\left(t_{0},\tilde{\tau}\right)$,
where $\tilde{\tau}=\tau_{1}+\tau_{2}+\dots+\tau_{k}$, by multiplication
of short-term transition matrices \cite{froyland2012finite}
\begin{align} \label{eq:Markov}
\mathbf{P}\left(t_{0},\tilde{\tau}\right)\approx\mathbf{P}\left(t_{0},\tau_{1}
\right)\mathbf{P}\left(t_{0}+\tau_{1},\tau_{2}\right)\cdot\dots \nonumber \\
\ldots \cdot 
\mathbf{P}\left(t_{0}+\tau_{1}+\dots+\tau_{\left(k-1\right)},\tau_{k}\right).
\end{align}
In this case, the information about the exact particle positions within each box is lost after every time step $\tau_{i},\ i=1,\dots,k$. This corresponds to intra-box mixing of the particles within each box at every time step mimicking the diffusive small-scale dynamics of real-world flows not resolved by fully deterministic flow models. The Markov approximation has been successfully applied to geophysical flows by various 
authors~\cite{froyland2014well,dellnitz2009seasonal, sergiacomi2015dominant}.

\section{Lagrangian flow networks}
\label{sec:network}

In the following, we interpret the transition matrix (Eq.~\ref{eq:ulam}) as the \emph{weight matrix} of a weighted and directed complex network, the \emph{Lagrangian flow network}. The box $B_i$ of the partition corresponds to the node $i$ of this network, and a link between nodes $i$ and $j$ is present if fluid is exchanged between boxes $B_i$ and $B_j$ during a time step $\tau$. Link weights are given by the entries of the transition matrix $P_{ij}\left(t_{0},\tau\right)$. 

The network's adjacency matrix $\mathbf{A}\left(t_{0},\tau\right)$ is thus defined as
\begin{equation}
A_{ij}\left(t_{0},\tau\right):=\begin{cases}
1, & \text{if}\ P_{ij}\left(t_{0},\tau\right)>0,\\
0, & \text{else}.
\end{cases}\label{eq:adjacency}
\end{equation}
\noindent
Based upon this adjacency matrix, a variety of complementary structural characteristics can be studied. In the following, we present a selection of these measures to be further used in the course of this study.

\subsection{Degree}

The \emph{out-degree} $k_i^{out}$ of a node $i$ is given by the number of edges pointing from node $i$ to any other node of the network. Although often excluded (e.g., in functional network representations of multivariate or spatio-temporal data fields~\cite{Radebach2013}), in this work we explicitly allow for self-loops, i.e., edges pointing from $i$ to $i$ (corresponding to the fraction of passive tracers staying within $i$ after a time step $\tau$). Conversely, the \emph{in-degree} $k_i^{in}$ counts the number of edges pointing to node $i$. Both quantities can be easily computed from the network's adjacency matrix $\mathbf{A}$ as
\begin{eqnarray}
k_i^{out} & = & \sum_{j=1}^{N} A_{ij},\label{eq:outdegree}\\
k_i^{in} & = & \sum_{j=1}^{N} A_{ji}.\label{eq:indegree}
\end{eqnarray}

The degree of a node in a Lagrangian flow network is directly related with the volume influenced by an intial perturbation in box $B_i$ after a time $\tau$. For a two-dimensional flow and a constant box size $m_B$, the affected area will be approximately $k_i^{out}\cdot m_B$. Therefore, the out-degree $k_i^{out}$ is a natural measure of the (nonlinear) spreading of particles, while the in-degree $k_i^{in}$ directly relates to the mixing of fluid of different origin inside box $B_i$. Based on these considerations, Ser-Giacomi \textit{et~al.}~\cite{sergiacomi2015flow} recently presented an argument that heuristically relates the degree with the FTLE.

\subsection{Eigenvector centrality}
 
A large body of recent work has utilized the eigenvectors of the transition matrix for studying geophysical flows~\cite{froyland2014well,vansebille2012origin}. Specifically, it has been demonstrated that the eigenvectors associated to eigenvalues close to 1 can be used to detect (almost) invariant sets in flow systems, i.e., in a set-oriented framework.

In this work, we consider the related measure of \emph{left eigenvector centrality} $x_i^{\;left}$ of a node $i$, which is defined based on the Lagrangian flow network's adjacency matrix $\mathbf{A}$ (instead of the transition matrix $\mathbf{P}$) as a solution of the eigenvalue problem
\begin{equation}
 \vec{x}^{\;left,\;T}\mathbf{A}=\lambda_{max}\vec{x}^{\;left,\;T},
\end{equation}
\noindent
where $\lambda_{max}$ is the largest eigenvalue of $\mathbf{A}$. Since all coefficients of $\mathbf{A}$ are non-negative, $\lambda_{max}$ and all coefficients of $\vec{x}^{\;left}$ (i.e., the left eigenvector centralities $x_i^{\;left}$) are also non-negative, and $\vec{x}^{\;left}$ is unique with these properties~\cite{Newman2010}. 

In a similar way, one may study the solution of the closely related eigenvalue problem
\begin{equation}
 \vec{x}^{\;right,\;T}\mathbf{A}^T=\lambda_{max}\vec{x}^{\;right,\;T},
\end{equation}
\noindent
where $x_i^{\;right}$ denotes the \emph{right eigenvector centrality} of node $i$. One easily convinces oneself that while the left eigenvector centrality (based on $\mathbf{A}$) is related to the time-forward dynamics, the right eigenvector centrality (based on $\mathbf{A}^T$) characterizes the time-backward dynamics.

The interpretation of eigenvector centrality is somewhat less straightforward than that of some other centrality measures used in complex network theory. Following the intuition that the degree gives the number of possible walks of length 1 from node $i$ to all other nodes, eigenvector centrality is related to number of possible walks of arbitrary length from node $i$ to all other nodes~\cite{bonacich1991simultaneous}. This heuristically explains why the patterns of eigenvector centrality and degree often exhibit a certain degree of similarity.

Regarding the two different numerical schemes for estimating the transition matrix with and without the adaptive condition (Eq.~\ref{eq:adap}) as described above, we may observe a minor difference in the resulting distribution of eigenvector centrality. Most notably, as will be shown in Section~\ref{sec:results}, the maxima in the field of (left) eigenvector centrality of $\mathbf{A}$ are more pronounced when using adaptive particle numbers per box, whereas the (right) eigenvector centrality of $\mathbf{A^T}$ remains largely unaffected by different choices.

\subsection{Closeness\label{sec:closeness}}

In order to determine which nodes of a flow network correspond to boxes with maximum dispersion, a natural idea is to look at such nodes that have direct influence on a large number of other nodes and thereby are able to spread particles over a wide part of the flow domain. The previously discussed degree is able to measure this property. Another vital idea is to consider how many steps it takes for particles starting at node $i$ to reach \emph{all} other nodes and associate strong dispersion with a node if the resulting \emph{paths} to reach all other nodes are on average especially short. The network property capturing this notion of dispersion is \emph{closeness centrality}.

A \emph{path} in a network is a sequence of nodes, such that every node is connected to its consecutive node by an edge. In directed networks the edges have to additionally point into the correct direction. The \emph{length }of a path is the number of nodes traversed along the path. This leads to the definition of a\emph{ shortest path} between two nodes, which is simply a path of minimum length. While there is a definite minimum path length between two nodes, there can be several distinct shortest paths. 

Let $l_{ij}$ denote the length of a shortest path between nodes $i$ and $j$. The expected graph distance from $i$ to a randomly chosen node $j$ is then given by
\begin{equation}
l_{i}:=\frac{1}{N-1}\sum_{j\neq i} l_{ij}.
\label{eq:farness}
\end{equation}
\noindent
The \emph{closeness centrality} $c_i$ of node $i$ is defined as the inverse average distance~\cite{Newman2010}
\begin{equation}
c_i:=\frac{1}{l_{i}}=\frac{N-1}{\sum_{j\neq i} l_{ij}}.
\label{eq:closeness}
\end{equation}
\noindent
If the studied network is not fully connected and nodes $i$ and $j$ belong to different components, it is convenient to define $l_{ij}:=N$, where $N$ is the total number of nodes in the network~\cite{Newman2010}. This value is always greater than the maximally possible shortest path length between two nodes. 

For directed networks, one has to distinguish \emph{out-closeness} and \emph{in-closeness} ($c_i^{out}$ and $c_i^{in}$, respectively), which just means that in Eq.~\eqref{eq:closeness} at each step only outgoing (incoming) edges are considered. As it is also the case for the degree, out-closeness relates to forward dynamics and in-closeness to backward dynamics. 

Note again that when we approximate actual particle trajectories by network paths, we disregard their precise location inside each box at every time step. In other words, we implicitly assume that our transition matrix exhibits the Markov property (Eq.~\ref{eq:Markov}).

\subsection{Path-based measures with cutoff}

For shortest path-based network characteristics like closeness (as well as, e.g., local efficiency or betweenness not further studied in this work), it is possible to introduce a cutoff $L$ and consider only those shortest paths in the network which have a length smaller than or equal to $L$. In this way we restrict our attention to processes that occur within a certain time interval, namely within $L$ time steps. Consequently, nodes $i$ and $j$ cannot be connected if the shortest path between them is of length greater than $L$. For closeness, this would imply $d_{ij}=N$. When the cutoff exceeds the maximum shortest path length, we get back the usual closeness definition. 

The introduction of cutoff closeness has several advantages: One the one hand, it significantly reduces the computational costs for computing closeness, since only paths up to a certain length have to be taken into account. Even more, while path-based measures without cutoff relate to the infinite-time behavior of the system, the cutoff allows to focus their sensitivity to the particular time-scales of interest.

As a special case, it is straightforward to show that the out-closeness with cutoff $L=1$ is a strictly monotonic transformation of the out-degree:
\begin{equation}
 c^{out,1}_i = \frac{N-1}{k^{out}_i+\left(N-1-k^{out}_i\right)\cdot N}=\frac{1}{N-k^{out}_i}.
 \label{eq:odeg-oc1}
\end{equation}
\noindent
For general values of $L$, it is possible to derive a similar expression. For this purpose, let $\kappa^L_i$ denote the number of nodes that can be reached from node $i$ after exactly $L$ steps (but \emph{not} at any time earlier). Then, we can express $c^{out,L}_i $ as
\begin{equation}
c^{out,L}_i = \frac{N-1}{\sum_{l=1}^{L}\kappa^l_i\cdot l + \left(N - 1 - \sum_{l=1}^{L}\kappa^l_i\right)\cdot N}. \label{eq:cclose_general}
\end{equation}
\noindent
Equation~(\ref{eq:cclose_general}) presents an intimate link between the cutoff-$L$ out-closeness and the ``generalized out-degrees''~$\kappa^L_i$. We note that $\kappa^L_i$ can be expressed exclusively by elements of the transition matrix $\mathbf{P}$ and its consecutive powers up to $\mathbf{P}^L$.

For $L > L_{\mathbf{P}}$, where $L_{\mathbf{P}}$ is the diameter of the network defined by the transition matrix $\mathbf{P}$, i.e., the maximum shortest path length over all pairs of nodes, we obtain:
\begin{equation}
c^{out,L}=c^{out,L_{\mathbf{P}}}=c^{out}= \frac{N-1}{\sum_{l=1}^{L_{\mathbf{P}}}\kappa^l_i\cdot l}.
\label{eq:cclose}
\end{equation}

In the context of our present work, we observe that for low values of $L$, $\kappa^L_i$ corresponds approximately to the normal out-degree of the matrix $\mathbf{P}^L$ as defined by Eq.~\eqref{eq:outdegree}. Recalling recent findings demonstrating a connection between the FTLE and the degree of (powers of) the corresponding transition matrix~\cite{sergiacomi2015flow}, one would expect a similar correspondence between the FTLE obtained with low integration times and out-closeness with low cutoffs for the vortex system. We will statistically evaluate this relationship for some example flow patterns in Section~\ref{sec:results}.

From Eq.~(\ref{eq:cclose}), one can see that out-closeness without cutoff essentially corresponds to an average of the (discretized) shortest possible travel times of flow between a box $B_i$ and all other boxes $B_j$ in the flow domain measured in units of $\tau$. Thus, out-closeness has a distinct temporal interpretation, indicating the time-scales at which a node is able to influence the rest of the network. In the context of geophysical flows, this property could be exploited for identifying regions from which particles diverge and spread very fast over the whole flow domain. In many situations, these region are of superior relevance, e.g., when studying the spread of non-reactive contaminants in the atmosphere or ocean.

Cutoff closeness retains the aforementioned temporal interpretation indicating fast dispersion, but additionally incorporates information on the part of the flow domain that can be influenced within $L$ time steps by effectively disconnecting pairs of nodes that are separated by more than $L$ steps.

\subsection{Treatment of time-dependent flows}

To compute local flow network characteristics as discussed above, we have implicitly assumed that the topology of the network given by the adjacency matrix $\mathbf{A}$ is constant during each time step. However, for time-dependent systems this is rarely the case. In general, $\mathbf{A}$ depends on the time step $\tau$ as well as on the initial time $t_{0}$. 

For the time-periodic example system studied in the following, we choose $\tau$ to match the systems' respective time-periodicity, so that $\mathbf{A}$ is in fact constant. In the more general case, $\mathbf{A}$ would vary with each time step, and we would have to deal with a sequence of distinct adjacency matrices $\mathbf{A}_{1},\dots,\mathbf{A}_{k},\dots$. If the number of time steps is finite, the theory of \emph{time-ordered graphs} allows to compute closeness for this time-dependent network topology~\cite{kim2012temporal}. To this end, however, a detailed exploration of this general case should remain a subject of future work.

\section{Driven vortex flow}\label{sec:example}

An omnipresent phenomenon in geophysical flows is the formation of vortices, such as cyclones, tornadoes, or ocean gyres. In the following, we study a generic model system that describes a chain of periodically driven vortices~\cite{tabeling1987instability,tabeling1990chaos,witt1999tracer,feudel2005intersections}.
In its simplified version studied by Feudel \textit{et~al.}~\cite{feudel2005intersections}, a thin layer of Lagrangian chaos is formed around a vortex embedded in a laminar shear flow. This chaotic layer contains regions of high dispersion and mixing, making the model especially useful for testing dispersion measures. 

Let us consider the stream function~\cite{feudel2005intersections}
\begin{equation}
\psi\left(x,y,t\right)=A\left(1+\varepsilon\sin\pi 
t\right)\sin y+B\sin 2x\sin y.\label{eq:vortex_sf}
\end{equation}
\noindent
Here, the first term generates a shear flow and the second term a chain of driven vortices. $\varepsilon$ is a constant that measures the strength of temporal modulation responsible for the emergence of chaotic particle motion in the flow. In accordance with previous studies, we set the values of the coefficients to $A=8.35$ and $B=-2.55$. Since we are not interested in the influence of varying modulation on the system, but want to obtain a general understanding of our network-based analysis method, we fix $\varepsilon=0.2$ - a setting also discussed by Feudel \textit{et~al.}~\cite{feudel2005intersections}. 

Considering $\psi\left(x,y,t\right)$ as a time-dependent Hamiltonian, we obtain the equations of motion for passive tracer particles as 
\begin{equation}\label{eq:hamiltonian}
\frac{dx}{dt}=\frac{\partial\psi}{\partial y},\quad \frac{dy}{dt}=-\frac{\partial\psi}{\partial 
x},
\end{equation}
\noindent
yielding
\begin{eqnarray}
\frac{dx}{dt} & = & A\left(1+\varepsilon\sin\pi t\right)\cos y+B
\sin 2x \cos y \nonumber \\
\frac{dy}{dt} & = & -2B\cos 2x \sin y \label{eq:vortex_coord}.
\end{eqnarray}

We note that the system is time-periodic with period $2$ and exhibits spatial periodicity in $x$-direction with period $\pi$ and in $y$-direction with
period $2\pi$. Furthermore, the $y$-nullclines are given by $\dot{y}\left(x,0,t\right)=\dot{y}\left(x,\pi,t\right)=0$ and are impassable barriers for trajectories for all $x$ and $t$. Therefore, we consider the phase space of the tracer dynamics restricted to the square $\left[0,\pi\right]\times\left[0,\pi\right]$. Accordingly, our numerical simulations are run with periodic boundary conditions in $x$-direction on this square. 

\begin{figure}
 \includegraphics[width=\columnwidth]{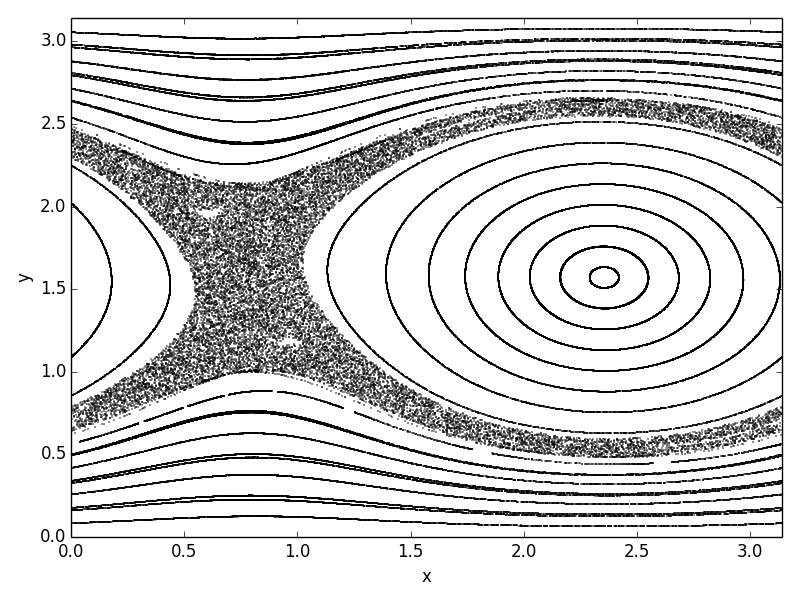}
 \caption{Stroboscopic map of the driven vortex flow (Eq.~\ref{eq:vortex_coord}) with $\varepsilon=0.2$. 48 particles were initialized along the $x=\pi/2$ and $x=3\pi/2$ axes and advected for 2000 time steps of length $\tau=2$ matching the natural period of the system. After every step, the location of each particle is plotted.}
 \label{fig:Stroboscopic-map}
\end{figure}

Figure~\ref{fig:Stroboscopic-map} shows a stroboscopic map of the system. We clearly discern the region of chaotic dynamics that separates the periodic region inside the vortex from the shear flow. Notably, the flow exhibits a saddle point at $\left(\frac{\pi}{4},\frac{\pi}{2}\right)$ and an elliptic fixed point at $\left(\frac{3\pi}{4},\frac{\pi}{2}\right)$. For $\varepsilon>0$, the invariant manifolds associated with the hyperbolic fixed point are stretched and folded, generating a chaotic saddle of mutual intersections between both manifolds which forms the backbone of the chaotic layer~\cite{feudel2005intersections}.

\section{\label{sec:results}Results}

For the driven vortex model described in detail in the previous section, we construct a transition matrix representation of the flow by partitioning the flow domain $\left[0,\pi\right]\times\left[0,\pi\right]$ into $200\times200$ boxes of equal size. We initialize 50 particles within every box at $t_0=0$ and advect them for a time step $\tau=2$ matching the period of the system. Then, we  adaptively increase the number of particles according to Eq.~\eqref{eq:adap}. In the following discussions, all network characteristics have been computed from the thus obtained transition matrix.

\subsection{Spatial patterns of network measures}

\subsubsection{Degree and eigenvector centrality}

\begin{figure*}
 \subfigure{
 \includegraphics[width=\columnwidth]{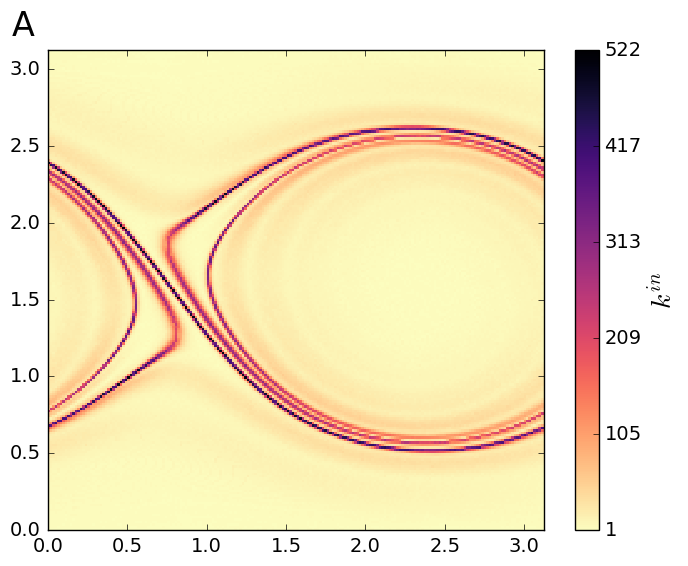}
  }
 \subfigure{
 \includegraphics[width=\columnwidth]{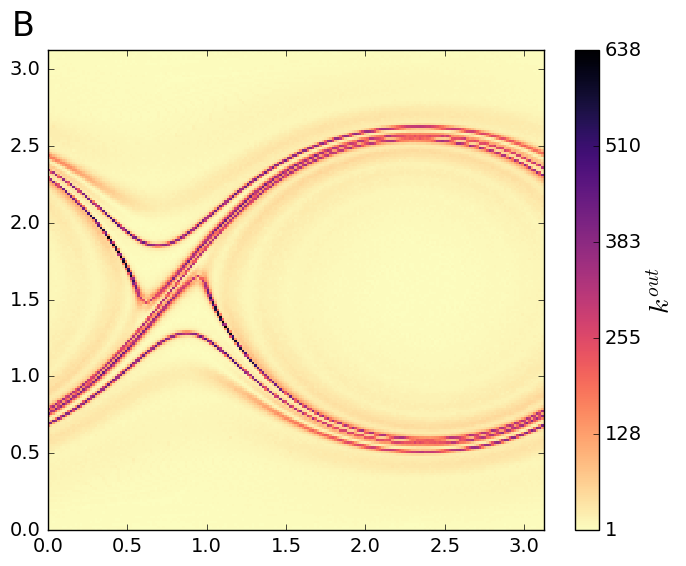}
  }
 \subfigure{
 \includegraphics[width=\columnwidth]{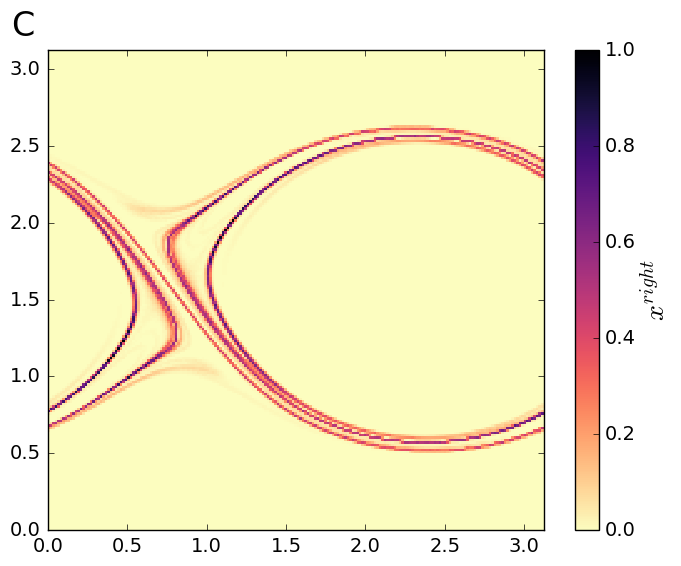}
  }
 \subfigure{
 \includegraphics[width=\columnwidth]{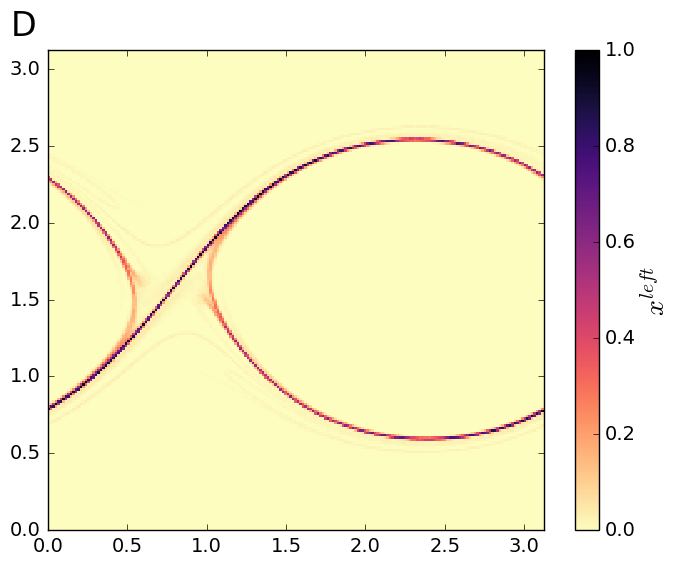}
  }
 \subfigure{
 \includegraphics[width=\columnwidth]{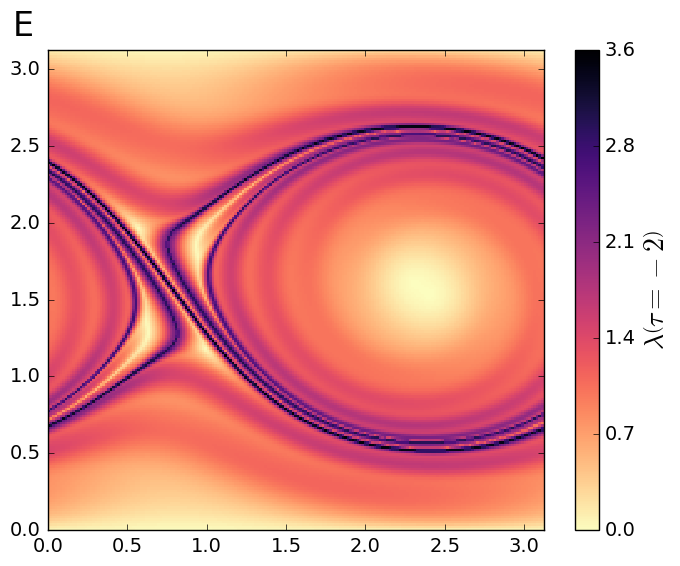}
  }
 \subfigure{
 \includegraphics[width=\columnwidth]{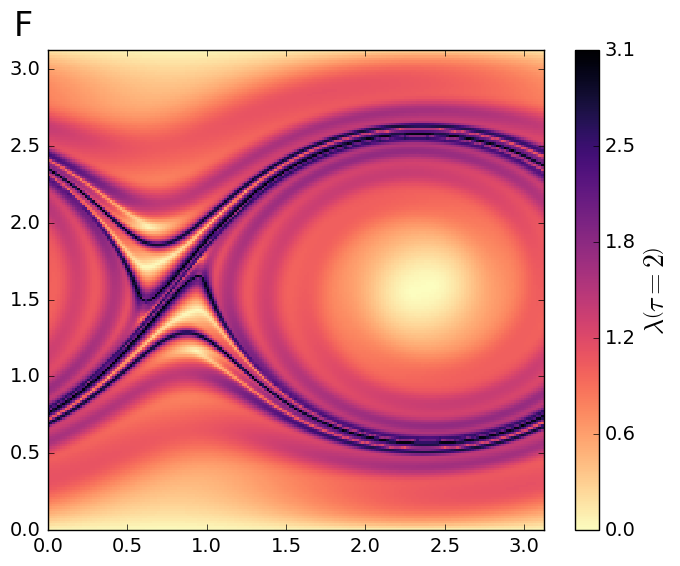}
  }
  \caption{(Color online) Time-backward (left) and time-forward measures (right) for the Lagrangian flow network of the driven vortex system: (A,B) in-/out-degree, (C,D) right/left eigenvector centrality and FTLE for (E) $\tau=-2$ and (F) $\tau=2$.}
  \label{fig:vortex-1}
\end{figure*} 

Figure~\ref{fig:vortex-1} displays the spatial patterns of degree and eigenvector centrality together with the associated FTLE fields (for $\tau=-2$ and $\tau=2$) for forward (right panels) and backward (left panels) dynamics. It can be clearly seen that out-degree and in-degree highlight the stable and unstable manifolds of the hyperbolic point, respectively. The ridges described by the local maxima of the degree fields are in very good correspondence with the positions of the invariant manifolds as studied by \citet{feudel2005intersections}, which is to be expected from their close relationship with the FTLE fields, the ridges of which are commonly used for approximating invariant manifolds \cite{Haller2002,shadden2005definition}. Note that as already emphasized in Section~\ref{sec:methods}, the out-degree takes on average higher values than in-degree in the considered sampling scheme.

The patterns of right and left eigenvector centralities closely resemble those of in- and out-degree, respectively, which underlines the conceptual relationship between eigenvector centrality and degree (the former being a straightforward extension of the latter) \cite{Newman2010}. However, we observe that the time-forward measures differ more from each other than their time-backward counterparts. As stated before, this is due to the adaptively chosen number of particles per box, which is optimized with respect to the out-degree (i.e., a measure related to the time-forward dynamics of the flow).

\subsubsection{Out-closeness with and without cutoff}

\begin{figure*}

 \subfigure{
 \includegraphics[width=\columnwidth]{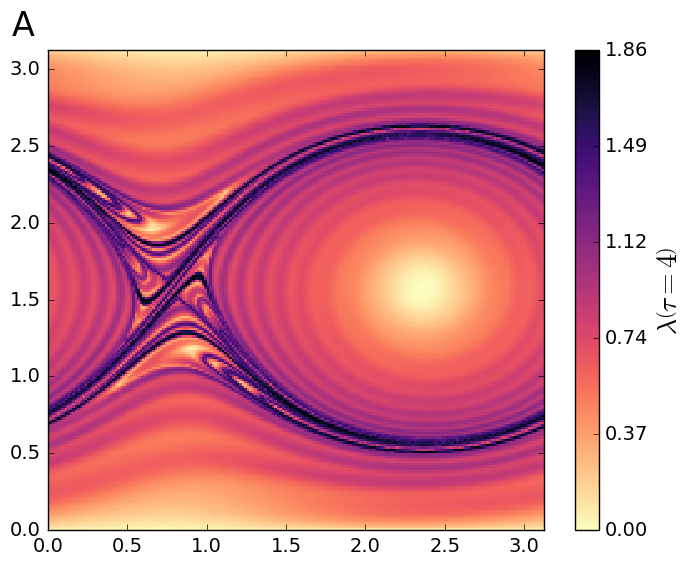}
  }
 \subfigure{
 \includegraphics[width=\columnwidth]{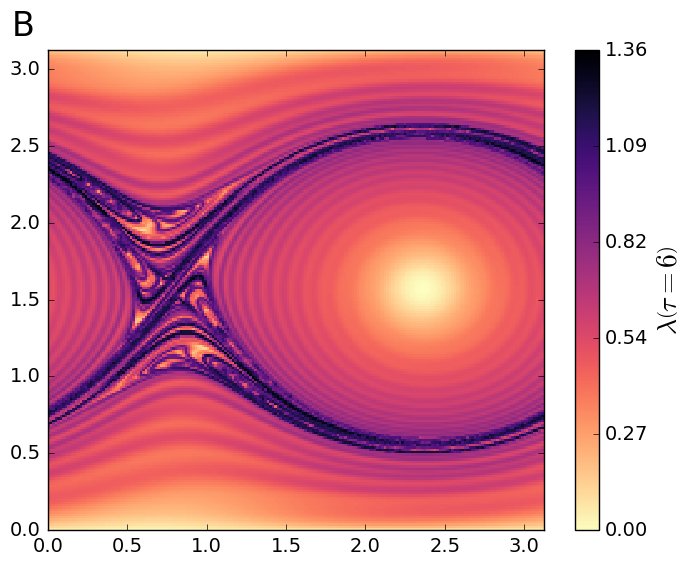}
  }
 \subfigure{
 \includegraphics[width=\columnwidth]{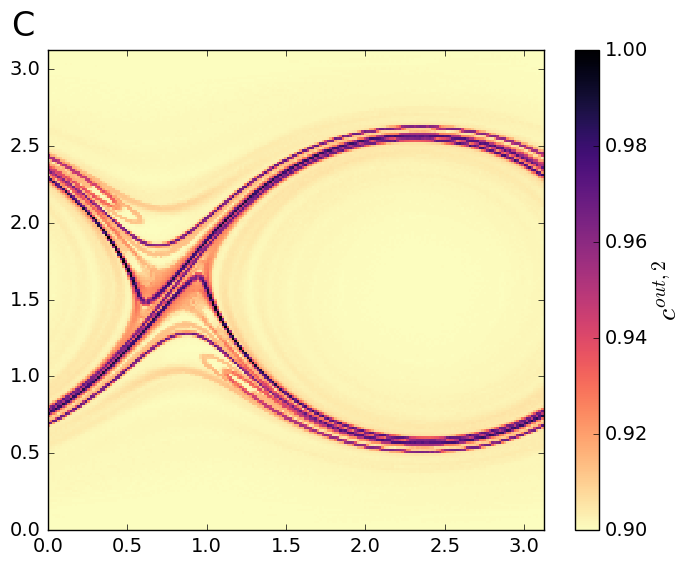}
  }
 \subfigure{
 \includegraphics[width=\columnwidth]{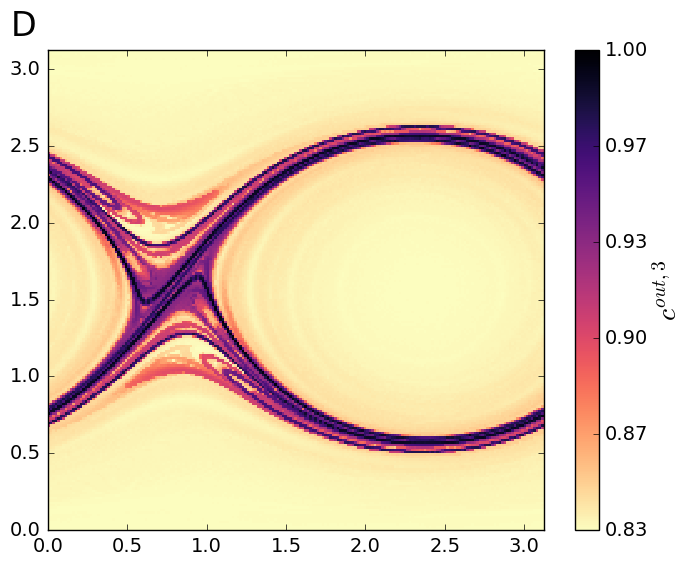}
  }
 \subfigure{
 \includegraphics[width=\columnwidth]{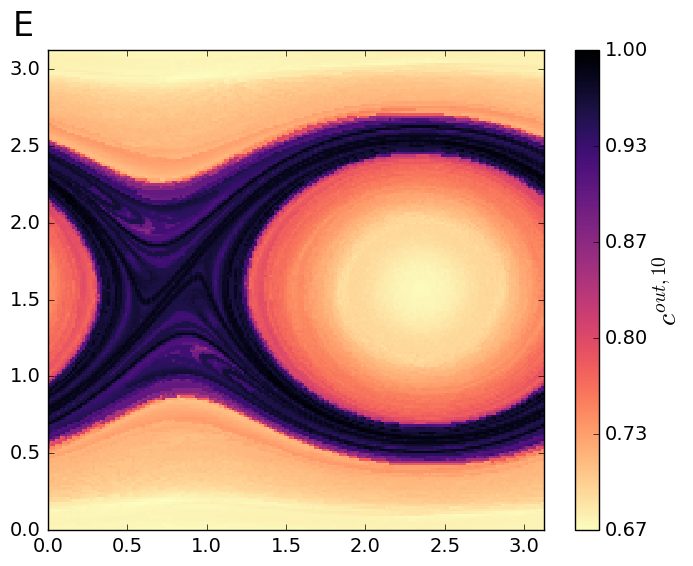}
  }
 \subfigure{
 \includegraphics[width=\columnwidth]{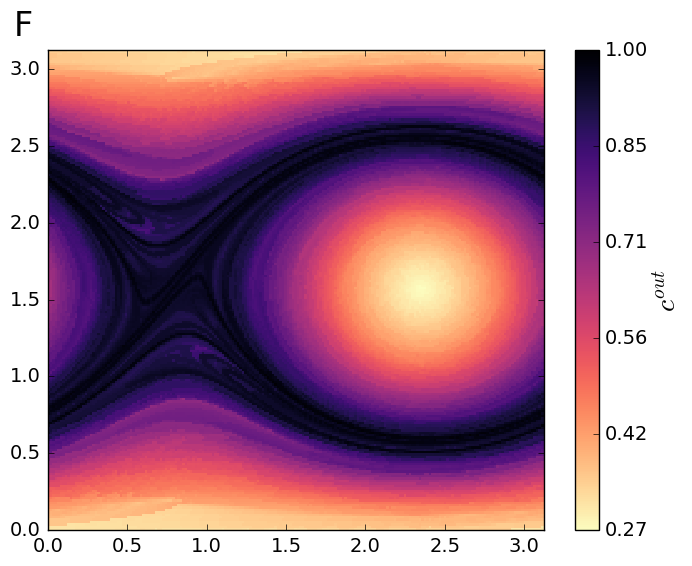}
  }
  \caption{(Color online) (A,B) Time-forward FTLE and (C-F) out-closeness with different cutoff values $L$ (normalized by their respective maximum values for the whole flow network) for the driven vortex system. We chose the integration time as (A) $\tau=4$ and (B) $\tau=6$, respectively. The considered cutoff values $L$ are (C) $L=2$, (D) 3, (E) 10 and (F) $\infty$ (i.e., no cutoff was used). Recall that $c^{out,1}$ corresponds to a monotonic transformation of the out-degree, which has been shown in Fig.~\ref{fig:vortex-1}B.}
\label{fig:vortex-2}  
\end{figure*}

By varying the cutoff value $L$ for the maximum path length in the closeness computation, we get direct access to the system's dispersion characteristics at different time-scales. Figure~\ref{fig:vortex-2} shows the spatial patterns of cutoff out-closeness for different cutoffs together with the forward FTLE fields with different integration steps $\tau$.

For small $L>1$ (Fig.~\ref{fig:vortex-2}C,D), cutoff out-closeness highlights the stable manifold of the hyperbolic point embedded in the chaotic layer, but also exhibits additional structures in comparison with the out-degree. Notably, we observe similar structures when considering the time-forward FTLE field with the corresponding integration time (Fig.~\ref{fig:vortex-2}A,B).

For larger cutoff values (Fig.~\ref{fig:vortex-2}E), we obtain a partitioning of the phase space into regions of turbulent and laminar dynamics, with high closeness values highlighting the chaotic layer. The two KAM tori embedded in this layer remain visible for every cutoff value as local minima of the closeness field, which is in agreement with the previously reported observation of local FTLE minima around these tori \cite{feudel2005intersections}. This means that even small regions of regular dynamics embedded in the chaotic layer can be identified by cutoff closeness.

Finally, increasing the cutoff $L$ even further (and finally removing it completely) results in a smoother transition between the dynamically different regimes. 

In summary, we find a striking similarity between the FTLE fields and cutoff closeness (as well as the out-degree being closely related with the cutoff-1 closeness). This suggests that instead of previously considered FTLEs or related measures based on transverse dynamics, flow network characteristics may be used for unveiling dynamically invariant (or almost invariant) structures in flows. In the following, we will present a more quantitative assessment of the corresponding similarities together with an inter-comparison between the different network properties.

\subsection{Statistical analysis}

For the out-degree, \citet{sergiacomi2015flow} provided a heuristic argument for establishing a theoretical relationship with the average FTLE in a given box. However, a corresponding formalization is much harder to achieve for the (cutoff) out-closeness studied in this work. Instead, we aim to provide further evidence for its relationship with the FTLE by means of a thorough statistical inter-comparison. In order to quantify a possibly nonlinear correspondence between the two fields, we consider two measures: (i) the area under curve ($AUC$)~\cite{fawcett2006introduction} associated with the receiver-operator characteristic (ROC) and (ii) symmetric uncertainty ($SU$)~\cite{press2007numerical}, a normalized version of the mutual information (MI).

\subsubsection{Receiver-operator characteristic}

Receiver-operator characteristic (ROC) analysis has been developed to quantify the performance of classification algorithms. In this context, for a given reference classification, the discriminatory threshold value of a potential classifier is continuously varied, and the true and false positive rates of the obtained classification are displayed as a closed curve. Under the ideal condition of a perfect classification, the area under this ROC curve (area under the curve, $AUC$) would take a value of 1. In turn, $AUC=0.5$ would correspond to the skills of a random classifier, where true and false positive rates are equal along the ROC curve.

In order to use ROC analysis for comparing the FTLE fields of our example system with the resulting spatial patterns of different flow network characteristics, we first have to define a classification target. In our case, we consider a binary version of the FTLE field distinguishing high and low values, the former being possibly related with the location of invariant manifolds. Specifically, we select a threshold $\lambda^*=\alpha\lambda_{\max}$ ($\lambda_{\max}$ being the maximum local FTLE value of the flow) with $\alpha\in\left(0,1\right)$ and define boxes with average FTLE value greater than $\lambda^*$ as ``positive'' and others as ``negative'' samples. By choosing $\alpha$ close to 1, we restrict our attention to those boxes that have the highest FTLE values, which are of special interest as argued below. In the next step, we consider the out-closeness value of every box as the output of a classifier, where high values are expected to correspond to a high probability of a positive sample.

In the literature on dispersion and mixing in geophysical flows, the relationship between ridges in the FTLE field and the presence of Lagrangian coherent structures and invariant manifolds has been intensively discussed~\cite{shadden2005definition, haller2011variational}. Here, a ridge represents a dense set of local maxima of the FTLE field in $d-1$ directions, where $d$ is the dimension of the system. Even though these ridges (i.e., locally maximal average FTLE values) do not necessarily coincide with the above defined positive samples, it is reasonable to assume that they are commonly contained in boxes with globally ``large'' average FTLE values. Since this relation to invariant manifolds and Lagrangian coherent structures is one of the most important properties of FTLE fields, it is of special interest how much information on local FTLE maxima can be obtained from the local properties of flow networks.

\begin{figure*}
 \subfigure{
 \includegraphics[width=\columnwidth]{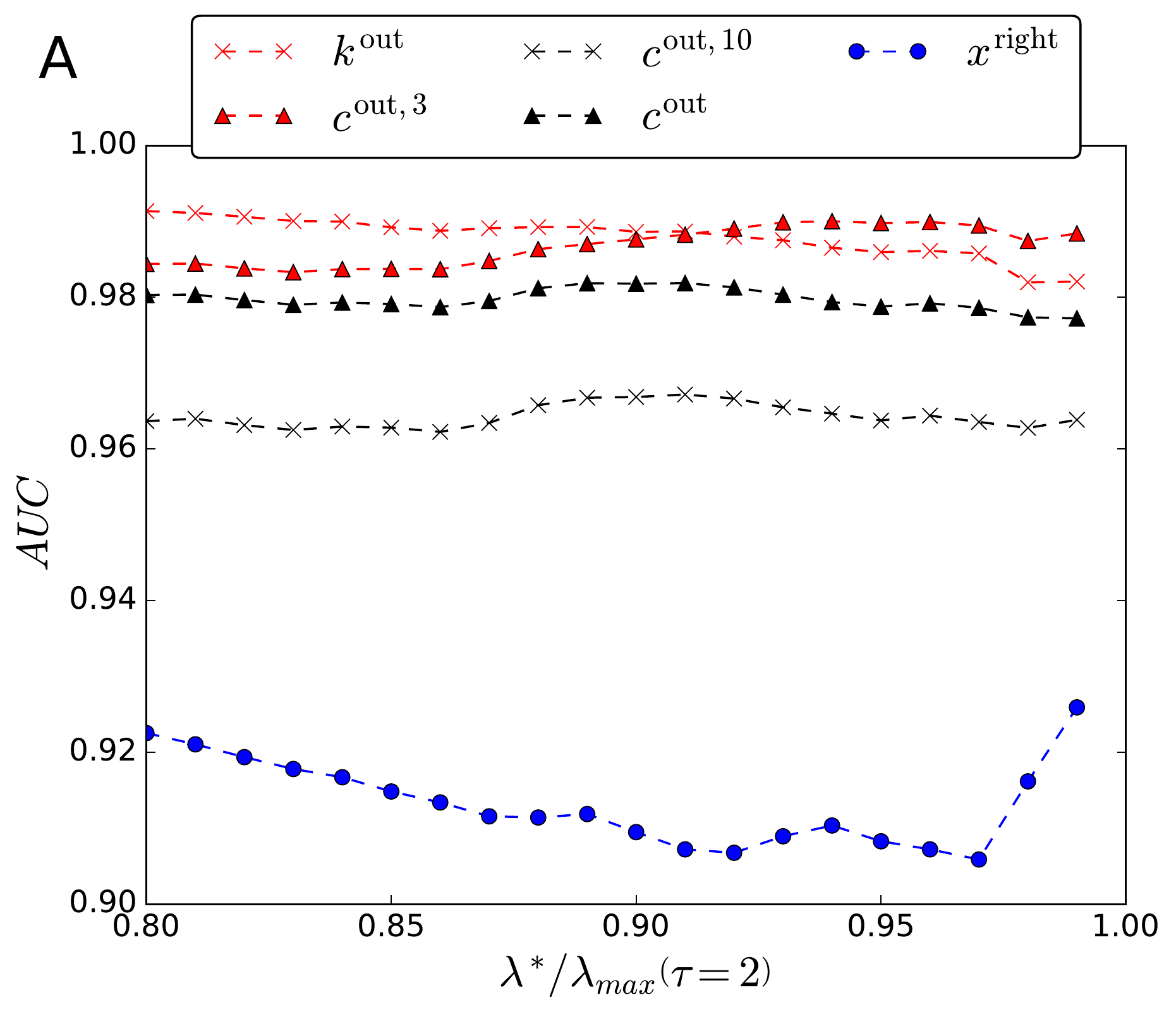}
  }
 \subfigure{
 \includegraphics[width=\columnwidth]{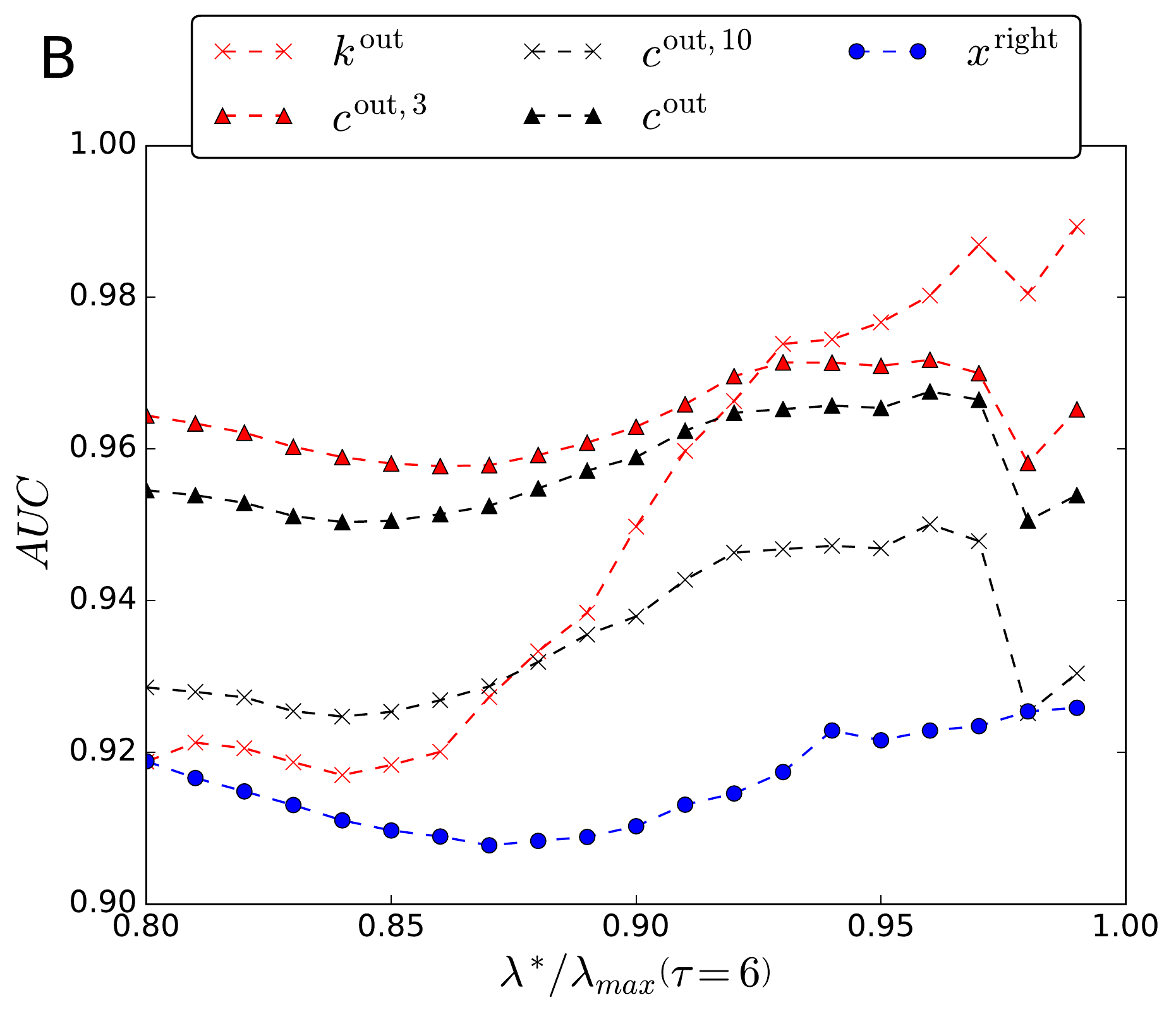}
  }
 \subfigure{
 \includegraphics[width=\columnwidth]{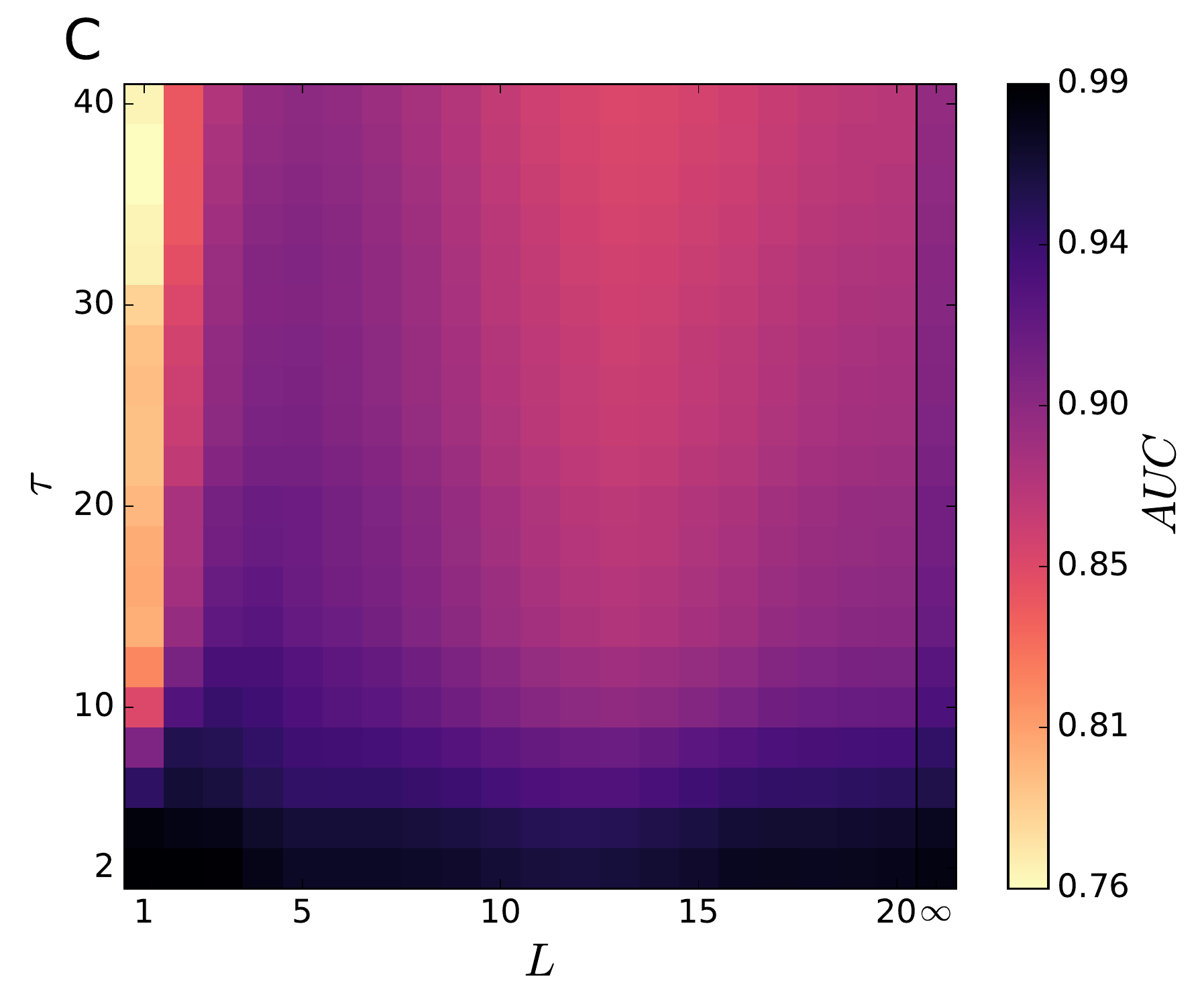}
  }
 \subfigure{
 \includegraphics[width=\columnwidth]{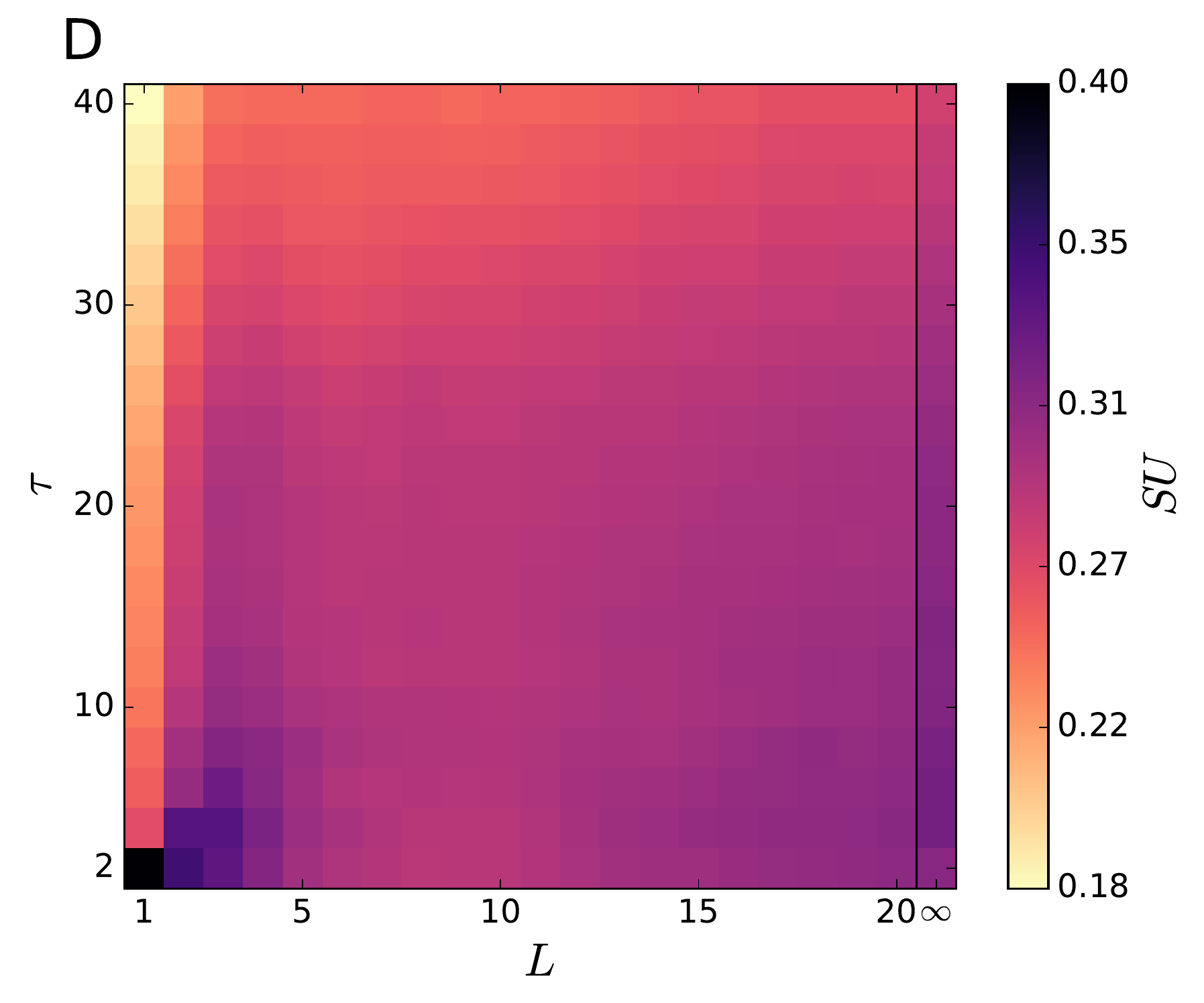}
  }
 \caption{(Color online) (A) $AUC$ for various network measures and different thresholds $\lambda^*$ to the FTLE field with $\tau=2$. (B) Same with $\tau=6$. (C) $AUC$ values for the cutoff-$L$ out-closeness obtained for an FTLE threshold of $\lambda^*/\lambda_{max}=0.90$ in dependence on the integration time $\tau$ and cutoff level $L$. (D) $SU$ values for the same setting as in (C). }
\label{fig:statistical-analysis}  
\end{figure*}

Figure~\ref{fig:statistical-analysis}A shows the obtained $AUC$ values for out-closeness with different cutoffs $L$ together with those obtained for out-degree and (right) eigenvector centrality when being compared to the FTLE field with an integration time of $\tau=2$. The curves are obtained by varying the threshold $\lambda^*$ over a reasonable range. All tested ``classifier fields'' yield $AUC$ scores higher than 0.9 and may thus be considered as providing essentially the same information as the FTLE in their maximum range. Among the considered flow network properties, out-degree (aka cutoff-1 out-closeness) and out-closeness with the comparatively low cutoff $L=3$ perform best with $AUC$ values of up to $0.99$ for certain $\lambda^*$ while the eigenvector centrality shows the least similarity with the FTLE field. In particular, when using a more restrictive threshold to the FTLE field (high $\lambda^*$), the cutoff-3 out-closeness outperforms the out-degree.

In Fig.~\ref{fig:statistical-analysis}B, the same analysis is repeated for FTLE with $\tau=6$. Here, over a wide range of $\lambda^*$, $c^{out,3}$ performs best. However, at very high $\lambda^*$ the out-degree reaches the highest $AUC$ values, which in turn performs much more poorly for lower $\lambda^*$. In general, the $AUC$ values appear to decrease with increasing integration time $\tau$, which reflects the loss of information on the longer-time evolution of particle trajectories in the coarse-grained network description. For both considered integration times $\tau$, it is remarkable that the ``normal'' out-closeness (without cutoffs) consistently performs better than out-closeness with cutoffs significantly larger than $L=\tau/2$ (i.e., the cutoff corresponding to the integration time).

In order to further investigate the latter phenomenon, we fixed a $\lambda^*=0.9\cdot\lambda_{max}$ and systematically compared FTLE and cutoff out-closeness for different values of $\tau$ and $L$. As Fig.~\ref{fig:statistical-analysis}C shows, the best results were obtained for combinations of $\tau=2$ and $\tau=4$ with either very low or very high values of $L$. The reason for this behavior is that most of the boxes covering the invariant manifold of the hyperbolic fixed point are already captured by the maxima of those low-$\tau$ FTLE fields, while the same invariant manifolds appear as maxima of all cutoff out-closeness fields.

\subsubsection{Mutual information and symmetric uncertainty}

Mutual information (MI) is a measure of general statistical interdependence between two random variables. In case of independence, the MI is 0, whereas it takes its maximum if one variable is a deterministic function of the other. In the latter case, the MI equals the Shannon entropy of the variables. In order to obtain a normalized measure, the symmetric uncertainty coefficient ($SU$) is defined as the ratio between MI and the mean of the marginal entropies. Hence, $SU$ takes the value 0 for independent variables and 1 if they are deterministic functions of each other. 

As an example, Figure~\ref{fig:statistical-analysis}D displays the obtained $SU$ values for different combinations of integration times $\tau$ for the FTLE estimations and cutoff levels $L$ for the cutoff out-closeness. We find that the correspondence between out-degree (aka cutoff-$1$ out-closeness) and the FTLE with $\tau=2$ is higher than for every other combination of $\tau$ and $L$. For $\tau$ and $L$ both taking low values, this particularly strong relationship is retained especially for $L=\tau/2$. However, if either $\tau$ or $L$ are further increased, the estimated values of $SU$ decrease and become almost independent of the specific combination with $SU\approx 0.3$ indicating again a non-trivial yet not necessarily very strong relationships between the FTLE and cutoff out-closeness fields.

Taken the results of ROC analysis and mutual information estimates together, we conclude that the similarity between FTLE fields with different integration times and cutoff out-closeness is especially large in the range of values close to the maximum FTLE. For short integration times and low cutoff levels, $c^{out,\tau/2}$ is commonly the best network-based classifier. In practical applications, we suggest that in case of doubts about the specific choice of $L$, the classical (no-cutoff) out-closeness should be chosen, since the mean values of both $AUC$ and $SU$ obtained for this measure and different integration times $\tau$ (Fig.~\ref{fig:statistical-analysis}C,D) were the highest among all cutoff levels $L$.

\section{\label{sec:discussion}Discussion and Conclusions}

Based on a discretized version of the transfer operator of a dynamical system obtained by Ulam's method, we have constructed Lagrangian transport networks as a tool for studying transport processes in flow systems. By making use of a Markov chain approximation, trajectories of passively advected particles can be approximated by paths on these networks. In order to provide insights into the properties of Lagrangian flow networks, we systematically studied the behavior of different established network characteristics, degree, eigenvector centrality and closeness, in the case of a simple model of a two-dimensional driven vortex flow. For the latter example, we demonstrated that degree and closeness with a variable cutoff $L$ to the maximum considered path length are efficient measures of dispersion and mixing, the spatial patterns of which strongly resemble those of the well-established FTLE fields. At the same time, evaluating these network characteristics comes on considerably lower computational costs, once the transition matrix has been obtained.

As a simplification, in this work we only studied the case of time-periodic systems, for which the transition probabilities between different parts of the flow domain are constant if the integration time is taken as an integer multiple of the period of the system. For the more complex problem of non-periodic time-dependent systems, the general idea of approximating particle trajectories by network paths could be considered as well, but requires further detailed investigations. First steps in this direction have been recently taken by Ser-Giacomi \emph{et al.}\cite{sergiacomi2014most, sergiacomi2015dominant}, making use of so-called time-ordered graphs~\cite{kim2012temporal}. 

Based on the obtained results, we suggest that Lagrangian flow network properties in general and (cutoff) out-closeness in particular provide useful tools for identifying dynamically (almost) invariant objects and Lagrangian coherent structures embedded in the flow. For this purpose, ridge detection algorithms commonly applied to FTLE fields can be used for the spatial patterns of the aforementioned network characteristics in a fully analogous way. In addition, we emphasize that the cutoff out-closeness has demonstrated a tendency of taking low values in even very small (of the order of box size) regions of regular dynamics, which might be used for identifying such regions directly from the transition matrix. For this purpose, also other network characteristics like betweenness or random-walk based versions of common path-based characteristics (e.g., random-walk closeness~\cite{Noh2004} or random-walk betweenness~\cite{Newman2005}) should be taken into account as candidate measures. We outline corresponding investigations, together with further applications to real-world flow systems, as subjects of future studies.

\section*{Acknowledgments}

This work has been financially supported by the German Federal Ministry for Education and Research (BMBF) within the framework of the BMBF Young Investigators Group CoSy-CC$^2$: Complex Systems Approaches to Understanding Causes and Consequences of Past, Present and Future Climate Change (grant no. 01LN1306A). The authors acknowledge stimulating discussions with Liubov Tupikina and Enrico Ser-Giacomi which put forward many of the ideas discussed in this work.

%

\bibliography{bibdat,bibreik}

\end{document}